\documentclass[onecolumn,aps,floatfix,pra,superscriptaddress,nofootinbib]{revtex4}
\usepackage{color}
\usepackage{graphicx}
\usepackage{bm}
\usepackage{amsmath}
\usepackage{amsbsy}
\usepackage{hyperref}
\usepackage{enumerate}
\usepackage{upgreek}
\usepackage[subnum]{cases}
\usepackage{dsfont}

\newcommand{\tr}{ \text{tr} }

\newcommand{\re}{ \text{Re} }

\newcommand{\pa}{ \partial }
\newcommand{\hb}{ \hbar }
\newcommand{\si}{ \sigma }
\newcommand{\om}{ \omega }

\newcommand{\ga}{ \gamma }
\newcommand{\la}{ \langle }
\newcommand{\ra}{ \rangle }

\newcommand{\al}{ \alpha }

\newcommand{\Ai}{ \text{Ai} }

\newcommand{\QM}{ \text{q} }

\begin{document}

\title{ Different theoretical aspects of the intrinsic decoherence in the Milburn formalism } 

\author{S. V. Mousavi}
\email{vmousavi@qom.ac.ir}
\affiliation{Department of Physics, University of Qom, Ghadir Blvd., Qom 371614-6611, Iran}
\author{S. Miret-Art\'es}
\email{s.miret@iff.csic.es}
\affiliation{Instituto de F\'isica Fundamental, Consejo Superior de Investigaciones Cient\'ificas, Serrano 123, 28006 Madrid, Spain}
\begin{abstract}

In this work, we consider different theoretical aspects and simple applications of the Milburn equation which is governed by a parameter controlling what is known as intrinsic decoherence. The main goal is to show some similarities also observed  when external decoherence is considered. Linear entropy, Ehrenfest relations, probability density current, the Wigner representation as well as the relation to a Lindbladian master equation are analyzed in terms of this intrinsic decoherence, leading to new insights on the Milburn dynamics. Interference of two wave packets, tunneling and the bouncing ball problem are also studied under this perspective. 

\end{abstract}

\maketitle

{\bf{Keywords}}: Intrinsic decoherence; Linear entropy; Ehrenfest relations; Arrival time; Wigner function; Tunneling; Bouncing ball.



\section{Introduction}

Quantum mechanics is a linear theory, and thus it admits coherent superposition of distinct physical states. 
This property is the basis of the measurement problem in this theory. However, this superposition principle does 
not act at the macroscopic scale, although nothing shows this in the formulation of the quantum mechanics. 
The problem of the standard quantum mechanics is to explain non-observability of macroscopic distinct states. 
Two routes have been tackled in the literature to overcome this problem. First, the role played by the environment. 
In this way, the whole system, the system of interest and the environment, is considered a 
closed system evolving on the standard Schr\"odinger equation. Then, by integrating out the environment
degrees of freedom, an equation known as the master equation is derived for the reduced density matrix of 
the system of interest. This formalism is known as environment-induced decoherence started by the work of 
Zeh \cite{Ze-FP-1970} and developed by Zurek \cite{Zu-RMP-2003}. Decoherence is important, not only from 
the foundational point of view but also in the realm of quantum information theory, since quantum 
information processing and quantum technologies depend on our ability in creation, maintenance and 
manipulation of non-classical superposed states \cite{Sc-PR-2019}.
In the second approach, the Schr\"odinger equation is itself modified by adding some additional terms which 
induce decoherence within the quantum mechanics itself. Among the various models have been proposed in 
this regard, the famous one is maybe the  model of Ghirardi, Rimini and Weber, the so-called GRW 
model \cite{GRW-PRD-1986}. In another approach, decoherence has been related to the gravity effects \cite{ElMoNa-PLB-1989}.

Over the years, several different paths have been taken to break up with the standard quantum mechanics; some of the well-known attempts include the postulation of extra nonlinear terms or stochastic terms in the Schr\"odinger equation; and gravitational effects invalidates this equation \cite{St-PTRSA-2012}. 
Several years ago, Milburn \cite{Mi-PRA-1991}  modified the von Neumann equation of the density operator 
in a way to include the decoherence, what is known as intrinsic decoherence. He postulated  that, at 
sufficiently short time steps, the system does not evolve  continuously under unitary evolution but  
by a stochastic sequence of identical unitary transformations.  This means that  a minimum time step in 
the universe is introduced. The inverse of this time step is the mean frequency of the unitary 
steps. The  system is then considered isolated and described by a state whose evolution is governed 
by a generalized von Neumann equation having this mean frequency as a parameter. 
For free particles, it can be seen that the Milburn equation is a Lindbladian master equation, 
where the only Lindblad operator is proportional to the square of linear momentum \cite{MoMi-JPA-2022}.
It is notable that in studying a Jaynes-Cummings model with phase damping, Kuang et. al \cite{Kuetal-PRA-1997} have introduced a master equation describing phase damping under the Markovian approximation which has the same form as the Milburn equation. 
If the frequency of time steps is sufficiently high, evolution in the experimental time-scales seems 
approximately continuous. In the zeroth order of the inverse of this frequency, the usual von Neumann equation is recovered. The first order 
correction yields decoherence in the energy eigenbasis. One aspect of this model is that constants of motion 
in the standard quantum mechanics approach remain constants of motion of the model and thus, stationary 
states remain stationary. Dynamics of open quantum systems has been considered within the Milburn 
framework \cite{BuKo-PRA-1998}. Furthermore, a Lorentz invariant model of intrinsic decoherence has 
been presented \cite{Mi-NJP-2006}.

The Milburn equation has attracted many attentions within the quantum information community in recent years. 
Effect of the intrinsic decoherence on quantum communications and various quantum correlations have been 
studied \cite{LiShZo-PLA-2010, JaNa-IJQI-2016, ZhZh-EPJD-2017, NaJa-IJTP-2019, AbMu-PS-2021, EsKhMaDa-OQE-2022, NaMuCh-PA-2022, OuChMa-IJTP-2023, KhMu-APB-2023, ChMa-APB-2023}. 
Decaying dynamics of a displaced harmonic oscillator \cite{UrMC-Pramana-2022} and in the mirror-field 
interaction \cite{UrMC-arxiv-2023} have also been studied  within this framework. 
In spite of these studies, only two works, to the best of our knowledge, have applied the Milburn equation in 
the coordinate representation: (i) Milburn himself investigated, in his original paper, the effect of the 
intrinsic decoherence on the interference fringes of the position probability density for a harmonic oscillator 
prepared in an initial superposition state of two coherent states \cite{Mi-PRA-1991} and (ii) this equation has been 
used by the same authors in the context of the quantum backflow \cite{MoMi-JPA-2022}. Thus, we think the 
Milburn equation deserves further attention by analyzing different theoretical aspects. In this paper, we are going 
to examine this equation in the coordinate representation.
Since, we sought to examine the first order correction to the customary von Neumann equation throughout the work, it is the first order approximation of the Milburn's equation that is used.

This work is organized as follows. Section II discusses some new aspects of the Milburn equation and the role played
by the intrinsic decoherence by analyzing linear entropy, Ehrenfest relations, the probability current density, arrival 
times, the Wigner representation, the relation to a Lindbladian master equation 
and transmission through a localized barrier. In section III, three simple problems are analyzed by comparing
the von Neumann and Milburn dynamics: interference of two Gaussian wave packet, tunneling through a 
rectangular barrier and the so-called bouncing ball problem. Finally, in the last section, some conclusions are 
reported.

\section{ Intrinsic decoherence in the Milburn formalism. Connection to the Wigner representation and the Lindbladian formalism}

The intrinsic decoherence within the Milburn formalism is based on the equation of motion of the density operator \cite{Mi-PRA-1991} given by 
\begin{eqnarray} \label{eq: Milburn0}
\frac{\pa \hat{\rho}}{\pa t} &=& \ga \left( e^{-i \hat{H} / \hb \ga} \hat{\rho} ~ e^{i \hat{H} / \hb \ga} - \hat{\rho} \right) .
\end{eqnarray}
In this equation of motion, $\hat{H}$ is the Hamiltonian of the system and $\ga$ a parameter which is the inverse 
of the minimum time step in the universe responsible for a stochastic sequence of identical unitary transformations 
at sufficiently short time steps.
Expansion of Eq. (\ref{eq: Milburn0}) up to first order in $\ga^{-1}$ reads as 
\begin{eqnarray} \label{eq: Milburn}
\frac{\pa \hat{\rho}}{\pa t} &=& - \frac{i}{\hb} [\hat{H}, \hat{\rho}] - \frac{1}{2\hb^2 \ga} [\hat{H}, [\hat{H}, \hat{\rho}] ] ,
\end{eqnarray}
which reduces to the well-known von Neumann equation in the limit $ \ga \rightarrow \infty $. 
In the energy representation, this equation can be written 
\begin{eqnarray} \label{eq: Mil_rhot_EEp}
\frac{\pa}{\pa t}\rho_{EE'}(t) &=& - \frac{i}{\hb} (E -E')\rho_{EE'}(t) - \frac{1}{2\hb^2 \ga} (E -E')^2 \rho_{EE'}(t) ,
\end{eqnarray}
$ \{ | E \ra \} $ being the energy eigenfunctions of the Hamiltonian $\hat H$ expressed as
\begin{eqnarray} 
\hat{H} | E \ra &=& E | E \ra  ,
\end{eqnarray}
where $ \rho_{EE'}(t) = \la E | \hat{\rho}(t) | E' \ra $ are the density matrix elements in the energy eigenbasis. Eq. \eqref{eq: Mil_rhot_EEp} 
is easily solved leading to 
\begin{eqnarray} \label{eq: rhot_EEp}
\rho_{EE'}(t) &=& \exp\left[- \frac{i}{\hb} (E -E')t - \frac{1}{2\hb^2 \ga} (E -E')^2 t \right] \rho_{EE'}(0)  .
\end{eqnarray}
Thus, for the density operator one has that
\begin{eqnarray} \label{eq: rhot_abs}
\hat{\rho}(t) &=& \sum_E \sum_{E'} \rho_{EE'}(t) |E\ra \la E'| \\
&=&  \sum_E \sum_{E'} e^{-i(E -E')t/\hb - (E -E')^2 t / (2\hb^2 \ga)  }\rho_{EE'}(0) |E\ra \la E'|  ,
\end{eqnarray}
which shows that if the initial state is an energy eigenstate then it does not evolve in time i.e., stationary states 
of the von Neumann equation remains so in the Milburn framework. 
Note that sums over $E$ and $E'$ turn into integrals when the energy spectrum is continuous.
From Eq. (\ref{eq: rhot_abs}), in the position representation one has that
\begin{eqnarray} \label{eq: rhot_EE'}
\rho(x, x',t) &=&  \sum_E \sum_{E'} e^{-i(E -E')t/\hb - (E -E')^2 t / (2\hb^2 \ga)  }\rho_{EE'}(0) u_E(x) u_{E'}^*(x') ,
\end{eqnarray}
where the corresponding diagonal elements of the density operator give us the probability density
\begin{eqnarray} \label{eq: rhot_xx}
P(x,t) &=& \sum_E \sum_{E'} e^{-i(E -E')t/\hb - (E -E')^2 t / (2\hb^2 \ga)  }\rho_{EE'}(0) u_E(x) u_{E'}^*(x)  ,
\end{eqnarray}
where $ u_E(x) = \la x | E \ra $. Now, if the initial state is considered to be a pure state
\begin{eqnarray} \label{eq: rho0}
\hat{\rho}(0) &=& | \psi_0 \ra \la \psi_0 | ,
\end{eqnarray}
then
\begin{eqnarray} \label{eq: prob_x}
P(x,t) &=& \sum_E \sum_{E'} e^{-i(E -E')t/\hb - (E -E')^2 t / (2\hb^2 \ga)  } C_E C^*_{E'} u_E(x) u_{E'}^*(x)  ,
\end{eqnarray}
where the expansion coefficients are given by
\begin{eqnarray} \label{eq: expan-coeff}
C_E &=& \la E | \psi_0 \ra = \int dx ~ u^*_E(x) \psi_0(x)  .
\end{eqnarray}
Due to the time-dependent damping exponential factor, the {\it long time behavior} of the probability density is 
given by
\begin{eqnarray} \label{eq: prob-lim}
\lim_{t \to \infty}P(x, t) & \approx & \sum_E |C_E|^2 |u_E(x)|^2 
\end{eqnarray}
which is independent on time. In this limit, non-diagonal elements of density matrix  decay 
exponentially. This is the way the intrinsic decoherence works in the Milburn framework. It is a gradual and continuous process
depending on the $\gamma^{-1}$ value. This behavior is the same observed for the external decoherence.

In the context of the standard quantum mechanics, when the Hamiltonian of a many body system 
has the additive form
\begin{eqnarray}
	\hat H &=& \sum_i \hat H_i   ,
\end{eqnarray}
a possible solution of the Schr\"odinger equation is written as a product of single particle states,
\begin{eqnarray}
	\Psi(x_1, x_2, \cdots x_N, t) &=& \Pi_i \psi_i(x_i, t)  ,
\end{eqnarray}
i.e., provided that the state is in the product form in an instant of time, this property is maintained over the time 
evolution. However, this is not valid  in the context of the Milburn framework.  
In fact, one can show that if the one-particle state $ \hat{\rho}_i $ evolves under the Milburn equation 
with the Hamiltonian $ \hat{H}_i $, then the two-particle separable state 
$ \hat{\rho} = \hat{\rho}_1 \otimes \hat{\rho}_2 $ is not evolved under the Milburn equation with the 
two-particle Hamiltonian $ \hat{H}_1 + \hat{H}_2 $ since
\begin{eqnarray}
	\frac{\pa}{\pa t} \hat{\rho}_1 \otimes \hat{\rho}_2 &=& 
	\frac{\pa \hat{\rho}_1}{\pa t} \otimes \hat{\rho}_2 + \hat{\rho}_1 \otimes \frac{\pa \hat{\rho}_2}{\pa t}
	\nonumber \\
	&=& \left( - \frac{i}{\hb} [\hat{H}_1, \hat{\rho}_1] - \frac{1}{2\hb^2 \ga} [\hat{H}_1, [\hat{H}_1, \hat{\rho}_1] ] \right) \otimes \hat{\rho}_2
	+ \hat{\rho}_1 \otimes \left( - \frac{i}{\hb} [\hat{H}_2, \hat{\rho}_2] - \frac{1}{2\hb^2 \ga} [\hat{H}_2, [\hat{H}_2, \hat{\rho}_2] ] \right)
	\nonumber \\
	&=& 
	- \frac{i}{\hb} [ \hat{H}_1 + \hat{H}_2, \hat{\rho}_1 \otimes \hat{\rho}_2] - \frac{1}{2\hb^2 \ga} [\hat{H}_1 + \hat{H}_2, [\hat{H}_1 + \hat{H}_2, \hat{\rho}_1 \otimes \hat{\rho}_2] ]
	+
	\frac{1}{\ga \hb^2} [\hat{H}_1,  \hat{\rho}_1] \otimes [\hat{H}_2,  \hat{\rho}_2]   .
\end{eqnarray}
The last term in the third line is responsible for this different behavior. Note that for convenience we have 
written $ \hat{H}_1 + \hat{H}_2 $ instead of 
$ \hat{H}_1 \otimes \hat{\mathds{1}}_2 + \hat{\mathds{1}}_1 \otimes \hat{H}_2 $; $ \hat{\mathds{1}}_i $ 
being the identity operator in the vector space of particle $i$.

\subsection{ Connection to the Wigner representation}

The Wigner function is defined as the {\it partial} Fourier transform of the density matrix in the $ (R, r) $ representation with respect to the relative coordinate $ r $ \cite{MoMi-Sym-2023}
\begin{eqnarray} \label{eq: Wigner}
W(R,u,t) &=& \frac{1}{2 \pi \hb} \int dr ~ e^{-i u r / \hb} \rho(R,r,t)   .
\end{eqnarray}
From Eq. \eqref{eq: mil-pos-Rr} one obtains an equation of motion for the Wigner function expressed as
\begin{eqnarray} 
\frac{\pa}{\pa t} W(R, u, t) &=& - \frac{u}{m}  \frac{\pa}{\pa R} W(R, u, t) 
+ \int du' ~ K(R, u'-u, t) ~ W(R, u',t)  \nonumber \\
&+& \frac{1}{2\pi \hb} \frac{-1}{2 \hb^2 \ga} \int du' ~ \int dr ~ e^{-iur/\hb} 
\left( \frac{ - \hb^2 }{m} \frac{\pa^2}{\pa R \pa r} + V(R+r/2) - V(R-r/2) \right)^2 
e^{i u' r / \hb} W(R, u',t)   ,
\nonumber \\
\label{eq: W-motion}
\end{eqnarray}
where the kernel of the first integral is defined as
\begin{eqnarray} \label{eq: kernel}
K(R, q,t) &=& \frac{1}{2 \pi \hb} \int dr ~ e^{i q r / \hb} \frac{ V(R/2+r) - V(R/2-r) }{\hb} .
\end{eqnarray}
Note that the first line of Eq. \eqref{eq: W-motion} is the equation of motion for the Wigner distribution function within the von Neumann formalism; the second line is the contribution due to the intrinsic decoherence. As an illustration, let us consider now two simple examples, the motion for a free particle and under a linear potential or constant force. 

For free particles, Eq. \eqref{eq: W-motion} is reduced to
\begin{eqnarray} \label{eq: Wig-eq-free}
\frac{\pa}{\pa t} W(x, p, t) + \frac{p}{m} \frac{\pa}{\pa x} W(x, p, t) - \frac{ p^2 }{ 2m^2 \ga } \frac{\pa^2}{\pa x^2} W(x, p, t) &=& 0   .
\end{eqnarray}
Apart from the last term which is responsible for the intrinsic decoherence, this equation has the same formal expression as that of the classical Liouville equation
\begin{eqnarray} \label{eq: Lio}
\frac{\pa}{\pa t} f(x, p, t) + \frac{p}{m} \frac{\pa}{\pa x} f(x, p, t) + \dot{p} \frac{\pa}{\pa p} f(x, p, t)  &=& 0 ,
\end{eqnarray}
where $f(x,p,t)$ is the phase space distribution function. Note that in this case $ \dot{p} = 0 $ and thus the last term of Eq. \eqref{eq: Lio} is zero. 
%
%
In the second example, the motion under  the presence of a linear potential
\begin{eqnarray} \label{eq: lin-pot}
V(x) = c_1 ~ x  ,
\end{eqnarray}
one obtains from Eq. \eqref{eq: W-motion}
\begin{eqnarray} \label{eq: Wig-eq-lin}
\frac{\pa}{\pa t} W(x, p, t) &+& \frac{p}{m} \frac{\pa}{\pa x} W(x, p, t) + c_1 \frac{\pa}{\pa p} W(x, p, t) 
\nonumber \\
&-& \frac{1}{\ga} \left( \frac{p^2}{2m^2} \frac{\pa^2}{\pa x^2} - \frac{c_1}{2 m} \frac{\pa}{\pa x} - c_1 \frac{p}{m}  \frac{\pa}{\pa x}  \frac{\pa}{\pa p} + \frac{c_1^2}{2}  \frac{\pa^2}{\pa p^2}
\right) W(x, p, t) = 0  .
\end{eqnarray}
which again, apart from the last term, one recovers the classical Liouville equation \eqref{eq: Lio}.

\subsection{ Connection to the Lindblad equation} \label{subse: mil-lind}

In the context of open quantum systems by taking into account the role played by the environment one obtains the Lindbladian master equation \cite{Sch-book-2007}
\begin{eqnarray} \label{eq: ld1}
	\frac{\pa}{\pa t} \hat{\rho}(t) &=& - \frac{i}{\hb} [\hat{H}', \hat{\rho}(t)] 
	- \frac{1}{2} \sum_i \kappa_i [\hat{L}_i, [\hat{L}_i, \hat{\rho}(t)]]   ,
\end{eqnarray}
for the evolution of the reduced density operator describing the system of interest where we have 
assumed Hermitian Lindblad operators $ \hat{L}_i $  and parameters $ \kappa_i >0 $. The so-called 
Lamb-shifted Hamiltonian $ \hat{H}' $ is not generally identical to the self Hamiltonian $ \hat{H} $ 
of the system containing additional terms caused by the interaction with the environment. 
For free particles with $ \hat{H}' = \hat{p}^2 / 2m $,  this master equation with only one Lindblad operator reads
\begin{eqnarray} \label{eq: ld2}
	\frac{\pa}{\pa t} \hat{\rho}(t) &=& - \frac{i}{\hb} [\frac{ \hat{p}^2 }{2m}, \hat{\rho}(t)] 
	- \frac{1}{2} \kappa [\hat{L}, [\hat{L}, \hat{\rho}(t)]] .
\end{eqnarray}
Comparison of Eq. \eqref{eq: ld2} with the Milburn equation 
\begin{eqnarray} \label{eq: Milburn1}
	\frac{\pa}{\pa t} \hat{\rho}(t) &=& - \frac{i}{\hb} [\frac{ \hat{p}^2 }{2m}, \hat{\rho}(t)] - \frac{1}{2\hb^2 \ga} [\frac{ \hat{p}^2 }{2m}, [\frac{ \hat{p}^2 }{2m}, \hat{\rho}(t)] ]   ,
\end{eqnarray}
for free particles shows that the Milburn equation \eqref{eq: Milburn1} has the same mathematical structure 
than the Lindbladian equation \eqref{eq: ld2} with a Lindblad operator proportional to $ \hat{p}^2 $. Let us
consider two simple cases.
If we consider $ \hat{L} = f(\hat{p}) $, $f$ being an arbitrary well-defined function, from Eq. (\ref{eq: ld2}) 
in the momentum representation we obtain
\begin{eqnarray} \label{eq: ld3}
	\frac{\pa}{\pa t} \rho(p, p',t) &=& \left\{ - \frac{i}{\hb} \frac{ p^2-p'^2 }{2m}
	- \frac{1}{2} \kappa (f(p)-f(p'))^2 \right\}\rho(p, p',t) ,
\end{eqnarray}
whose solution reads as
\begin{eqnarray} \label{eq: solld3}
	\rho(p, p', t) &=& \exp \left[ \left( - \frac{i}{\hb} \frac{p^2-p'^2}{2m} - \frac{1}{2} \kappa (f(p)-f(p'))^2 \right) t \right] \rho(p, p', 0) ,
\end{eqnarray}
revealing the probability density $ \rho(p, p, t) $ is time independent. 
%
%
%
%
Now, with the choice $ \hat{L} = \hat{x} $, Eq. (\ref{eq: ld2}) can be recast as
\begin{eqnarray} \label{eq: ld_pos-x}
	\frac{\pa}{\pa t} \rho(x, x', t) &=& 
	\left[ \frac{i \hb }{2m} \left( \frac{\pa^2}{\pa x^2} - \frac{\pa^2}{\pa x'^2} \right)
	- \frac{1}{2} \kappa (x-x')^2 \right]
	\rho(x, x', t)   ,
\end{eqnarray}
in the coordinate representation being the environmental scattering \cite{Sch-book-2007} equation of motion. 
Again, comparison of the high temperature limit of the well-known Caldeira-Leggett master equation 
\begin{eqnarray} \label{eq: CL eq}
	\frac{\pa}{\pa t} \rho(x, x', t) &=& \left[\frac{i \hb}{2m} \left( \frac{\pa^2}{\pa x^2} - \frac{\pa^2}{\pa x'^2} \right) - \lambda (x-x') \left( \frac{\pa}{\pa x} - \frac{\pa}{\pa x'} \right)
	- \frac{D}{\hb^2} (x-x')^2 \right] \rho(x, x', t) 
\end{eqnarray}
with Eq. (\ref{eq: ld_pos-x}) reveals that the former is Lindbladian only in the negligible dissipation limit i.e, 
when the second term can be neglected. Note that $\lambda$ is the relaxation rate or damping constant 
and $ D = 2 m \lambda k_B T_e $ plays the role of the diffusion coefficient ($k_B$ is the Boltzmann constant 
and $T_e$ is the temperature of the environment).

Finally, when comparing Eq. \eqref{eq: Milburn} to Eq.\eqref{eq: ld1}, one can see that the Milburn equation is just a master equation with the Hamiltonian as the only Lindblad operator.

\section{ First order approximation of Milburn's equation. Some analytical results }  

By using the first order approximation of Milburn's equation \eqref{eq: Milburn}, in this section we analyze 
the effect of the intrinsic decoherence on various important theoretical issues such as the linear entropy, 
Ehrenfest relations, arrival time distribution and transmission probability through a localized barrier 
providing some  analytical results.

\subsection{ Linear entropy }

Once it is well established this gradual intrinsic decoherence process, a standard quantity used in this context is the 
so-called linear entropy \cite{DoTaPaWa-PRA-1999}
\begin{eqnarray} \label{eq: pu_def}
S_L(t) &=& 1 - \tr( \hat{\rho}(t)^2 )   
\end{eqnarray}
being a measure of the mixedness of the state which is easier to calculate than the von 
Neumann entropy. In other words, the second term of Eq. \eqref{eq: pu_def} quantifies the purity of 
the state. Change of mixedness is a result of the non-unitary part of the evolution equation.
There are other applications beyond its main application as a measure of mixedness of the quantum state. For instance, in the field of quantum information theory, it has been used as a measure of quantum entanglement \cite{BuBoBe-PRA-2007}.

Since the trace operation is a basis independent quantity,  from Eq. (\ref{eq: Milburn0}), and writing the 
density matrix elements in the energy representation as $ \rho(E, E', t) = \la E | \hat{\rho}(t) | E' \ra $, one has that
\begin{eqnarray} \label{eq: denmat-en}
\rho_{EE'}(t) &=& \rho_{EE'}(0) ~ \exp\left[\ga t \left( e^{- i (E-E')/\hb \ga} - 1 \right) \right]  ,
\end{eqnarray}
and the linear entropy is expressed as 
\begin{eqnarray} \label{eq: pu}
S_L(t) &=& 1 - \sum_E ~  \la E | \hat{\rho}(t)^2 | E \ra 
= 1 - \sum_E \sum_{E'} ~  \la E | \hat{\rho}(t)| E' \ra \la E' | \hat{\rho}(t) | E \ra  \nonumber\\
&=& 1 - \sum_E \sum_{E'} ~  \la E | \hat{\rho}(0)| E' \ra \la E' | \hat{\rho}(0) | E \ra
\exp \left[ - 4 \ga t \sin^2 \left( \frac{E-E'}{2 \hb \ga}  \right) \right]   ,
\end{eqnarray}
where in the second line of this equation we have used Eq. (\ref{eq: denmat-en}). Thus, linear entropy decays also exponentially in time. 
Its short linear time behavior is then given by
\begin{eqnarray} \label{eq: pu_shorttime}
S_L(t) & \approx & S_L(0) + 4 \ga t \sum_E \sum_{E'} ~  | \la E | \hat{\rho}(0)| E' \ra |^2 \sin^2 \left( \frac{E-E'}{2 \hb \ga}  \right)   .
\end{eqnarray}
Here, we have used the Hermiticity property of the density operator. 
This linear behaviour in time is a typical feature of master equations (Markov approximation)
\cite{Joosetal-book-2003}. Note that as we mentioned in subsection \ref{subse: mil-lind}, 
the Milburn equation is a master equation with a single Lindblad operator taken to be the Hamiltonian 
of the system.  
The other interesting limit of linear entropy is its behavior when $\ga \rightarrow \infty$, i.e., the limit of the standard 
quantum mechanics. In this limit, to  first order in $\ga^{-1}$, one obtains from Eq. (\ref{eq: pu})
\begin{eqnarray} \label{eq: pu_1st}
S_L(t) & \approx & S_L(0) + \frac{t}{\ga \hb^2} \tr ( [\hat{H}, \hat{\rho}(0)]^2 )  .
\end{eqnarray}
Note that the operator $ [\hat{H}, \hat{\rho}(0)] $ is anti-Hermitian and thus its eigenvalues are imaginary. 
Thus, since the second term in Eq. (\ref{eq: pu_1st}) is negative, the intrinsic decoherence reduces 
entropy following a linear behavior in time.

\subsection{ Ehrenfest relations }

In the standard quantum mechanics, the Ehrenfest relations are
\begin{numcases}~
\frac{d}{dt} \la \hat{x} \ra = \frac{ \la \hat{p} \ra }{m} \label{eq: Eh1} \\
\frac{d}{dt} \la \hat{p} \ra = - \left \la \frac{\pa V}{\pa \hat{x}} \right \ra  .
\end{numcases}
It is instructive to see the effect of the intrinsic decoherence on these relations.
From Eq. (\ref{eq: Milburn0}), one obtains
\begin{eqnarray}
\frac{d}{dt} \la \hat{A} \ra_t &=& 
\ga \left \la e^{i \hat{H} / \hb \ga} \hat{A} e^{ - i \hat{H} / \hb \ga}  - \hat{A} \right \ra
\end{eqnarray}
for the rate of change of the expectation value of an observable $ \hat{A} $. To first order in $ \ga^{-1} $, 
one has that
\begin{eqnarray}
\frac{d}{dt} \la \hat{A} \ra_t & \approx & \frac{i}{\hb} \la [ \hat{H}, \hat{A} ] \ra_t - \frac{1}{2\hb^2 \ga} \la [\hat{H}, [ \hat{H}, \hat{A} ]] \ra_t  .
\end{eqnarray}
The rates of change of the expectation values of position and momentum operators are then
\begin{eqnarray}
\frac{d}{dt} \la \hat{x} \ra_t & \approx & \frac{ \la \hat{p} \ra_t }{m} - \frac{1}{2 m \ga} \left \la \frac{\pa V}{\pa \hat{x}} \right \ra_t \label{eq: pos_exp} \\
\frac{d}{dt} \la \hat{p} \ra_t & \approx & - \left \la \frac{\pa V}{\pa \hat{x}} \right \ra_t
- \frac{1}{4 m \ga} \left \la \left\{ \hat{p}, \frac{\pa^2 V}{\pa \hat{x}^2} \right \}\right \ra_t  ,
\label{eq: mom_exp}
\end{eqnarray}
where $ \{\hat{A}, \hat{B}\} = \hat{A}\hat{B} + \hat{B}\hat{A} $ is the anti-commutator of observables $ \hat{A} $ 
and  $ \hat{B} $. It is assumed here that the interaction potential is only a function of the position operator i.e., 
$ V = V(\hat{x}) $. These relations show that only for non-interacting (free) particles the first Ehrenfest relation 
(\ref{eq: pos_exp}) has the same formal expression as in the standard QM while for potentials, at most linear 
in the position operator, the second Ehrenfest relation (\ref{eq: mom_exp}) also does.
Then, one concludes that, only for free particles, $ \la \hat{x} \ra_t $ follows a classical trajectory. In fact, the
intrinsic decoherence does not affect the first moments $ \la \hat{x} \ra_t $ and $ \la \hat{p} \ra_t $ \citep{Mi-PRA-1991}.

As an illustration, let us consider the motion in a uniform gravity field for which $ V(\hat{x}) = mg \hat{x} $. 
In this case, solutions of Eqs. (\ref{eq: pos_exp}) and (\ref{eq: mom_exp}) yield
\begin{numcases}~ 
\la \hat{x} \ra_t \approx \la \hat{x} \ra_0 + \left( \frac{\la \hat{p} \ra_0}{m} - \frac{g}{2\ga} \right) t - \frac{1}{2} gt^2   ,
 \label{eq: pos_exp-linpot} \\
\la \hat{p} \ra_t \approx \la \hat{p} \ra_0 - m g t    . \label{eq: mom_exp-linpot} 
\end{numcases}
%
Note that while the expectation value of momentum is not affected by the intrinsic decoherence, the rate of change 
of the expectation value of position is reduced by the amount of $g/2\ga$. 
In the upward motion of a projectile, the peak height is 
$ \dfrac{ \la \hat{p} \ra_0^2 }{2 m^2 g} \left( 1 - \frac{m g}{\ga \la \hat{p} \ra_0} \right) $. 
Note that in order to reach this height, one needs $ \la \hat{p} \ra_0 > m g / \ga $.
If a harmonic potential $ V(\hat{x}) = m \om^2 \hat{x}^2 / 2 $ is now considered, solutions of 
Eqs. (\ref{eq: pos_exp}) and (\ref{eq: mom_exp}) are given by
\begin{numcases}~ 
\la \hat{x} \ra_t \approx e^{-\om^2 t / 2\ga} \left( \la \hat{x} \ra_0 \cos(\om t) + \frac{\la \hat{p} \ra_0}{m \om} \sin(\om t) \right)   ,
 \label{eq: pos_exp-har} \\
\la \hat{p} \ra_t \approx e^{-\om^2 t / 2\ga} ( - m \om \la \hat{x} \ra_0 \sin(\om t) + \la \hat{p} \ra_0 \cos(\om t) )   .\label{eq: mom_exp-har} 
\end{numcases}
Although the appearance of these equations suggests that the coefficient $ \ga $ acts like a friction
coefficient, this perception is misleading. Since the Milburn correction cannot change the energy, $ \ga $ cannot be the friction coefficient. 
The decay of the first order moments is not due to friction but to phase diffusion \cite{Mi-PRA-1991, Sa-PRA-1989, BaStPe-OC-1989} arising from the second term in Eq. \eqref{eq: Milburn}.

\subsection{ First moments of the arrival time distribution }

The Milburn equation \eqref{eq: Milburn} in the position representation reads as
\begin{eqnarray} \label{eq: mil-pos}
\frac{\pa}{\pa t} \rho(x, x', t) &=& 
- \frac{i}{\hb} \left[ \frac{ - \hb^2 }{2m} ( \pa_x^2  - \pa_{x'}^2 ) + V(x) - V(x') \right] \rho(x, x', t)
\nonumber \\
&-&
 \frac{1}{2\hb^2 \ga} 
\left[ \frac{ - \hb^2 }{2m} ( \pa_x^2  - \pa_{x'}^2 ) + V(x) - V(x') \right]^2 \rho(x, x', t)   .
\end{eqnarray}
This equation in the center of mass and relative coordinates, defined by $ R = (x+x') / 2 $ and $ r = x - x' $, is 
written as
\begin{eqnarray} \label{eq: mil-pos-Rr}
\frac{\pa}{\pa t} \rho(R, r, t) &=& 
- \frac{i}{\hb} \left[ \frac{ - \hb^2 }{m} \frac{\pa^2}{\pa R \pa r} + V(R+r/2) - V(R-r/2) \right] \rho(R, r, t)
\nonumber \\
&-&
 \frac{1}{2\hb^2 \ga} 
\left[ \frac{ - \hb^2 }{m} \frac{\pa^2}{\pa R \pa r} + V(R+r/2) - V(R-r/2) \right]^2 \rho(R, r, t)   ,
\end{eqnarray}
which in the absence of interactions i.e., for free particles, it can be written as a continuity equation
\begin{eqnarray} \label{eq: mil3}
\frac{\pa}{\pa t} \rho(R, r, t) + \frac{\pa}{\pa R} j(R, r, t)  &=& 0   ,
\end{eqnarray}
where we have introduced the probability current density matrix $ j(R, r, t) $ as
\begin{eqnarray} \label{eq: mil-curmat}
j(R, r, t)  &=& \left( - \frac{ i \hb}{m} + \frac{ \hb^2 }{2 \ga m^2} \frac{\pa^2}{\pa r \pa R}\right) \frac{\pa}{\pa r} \rho(R, r, t)  .
\end{eqnarray}
On the other hand, the density matrix in the position representation can be seen as the inverse Fourier transform 
of its representation in the momentum space,
\begin{eqnarray} \label{eq: rho_pos_mil}
\rho(R, r, t) &=& \frac{1}{2\pi \hb} \int_{-\infty}^{\infty}dp  \int_{-\infty}^{\infty}dp'
e^{i(p-p')R/\hb} e^{i(p+p')r/2\hb} \la p| \hat{\rho}(t) | p' \ra  ,
\end{eqnarray}
where, from the solution of Eq. \eqref{eq: Milburn} for free particles, we have that the matrix elements in this
representation are written as 
\begin{eqnarray} \label{eq: mil_sol}
\la p| \hat{\rho}(t) | p' \ra &=& \exp \left[ - \frac{i}{\hb} \frac{p^2-p'^2}{2m} t - \frac{1}{2\hb^2 \ga} \left( \frac{p^2-p'^2}{2m} \right)^2  t \right] \la p| \hat{\rho}(0) | p' \ra   ,
\end{eqnarray}
where the standard result is again reached when $\gamma^{-1} \rightarrow 0$.


By substituting now Eq. (\ref{eq: mil_sol}) into Eq. (\ref{eq: mil-curmat}), the diagonal elements 
of the probability current density are given by
\begin{eqnarray} \label{eq: mil-jd}
J(x, t)  &=& \frac{1}{4\pi m \hb} \int_{-\infty}^{\infty}dp  \int_{-\infty}^{\infty}dp' e^{i (p-p')x/\hb }
(p+p') \left( 1 - i \frac{p^2-p'^2}{4\hb m \ga} \right) \rho(p, p', t)   .
\end{eqnarray}
Note that, since the density operator is Hermitian, one has that 
$ \la p| \hat{\rho}(t) | p' \ra^* = \la p' | \hat{\rho}(t) | p \ra $ which, as expected, $ J(x, t) $ is real.

As a particular case, let us see if the product of states in the context of the standard quantum mechanics  is
kept in this formalism. If the wave function of a system is initially factorized as follows
\begin{eqnarray}
	\psi(\mathbf{r}) &=& \psi_1(x)\psi_2(y)\psi_3(z)  ,
\end{eqnarray}
one sees that it will be preserved along time when the potential is additive such as
\begin{eqnarray}
	V(\mathbf{r}) &=& V_1(x) + V_2(y) + V_3(z), 
\end{eqnarray}
(including $ V = 0 $). However, this result is not longer valid in the context of the Milburn equation. If one considers 
free propagation, Eq. (\ref{eq: mil_sol}) explicitly shows, even if the initial state is factorized as
\begin{eqnarray}
	\rho(\mathbf{p}, \mathbf{p'}, 0) &=& \rho_1(p_x, p'_x, 0) \rho_2(p_y, p'_y, 0) \rho_3(p_z, p'_z, 0)  ,
\end{eqnarray}
that the exponential term $ e^{ - [ (\mathbf{p}^2-\mathbf{p'}^2 ) / 2m ]^2  t/2\hb^2 \ga } $ appears  in 
Eq. (\ref{eq: mil_sol}) and this factorization is not maintained at later times.

Within the framework of the Bohmian mechanics, it is proved that the {\it ideal} (measurement-independent) 
arrival time distribution at the detector location $ x_d $ is uniquely given by the modulus of the probability 
current density when starting from the continuity equation \cite{Leavens}
\begin{eqnarray} \label{eq: ardis*}
\Pi(x_d, t) &=& \frac{ | J(x_d, t) |  }{ \int_0^{\infty}dt' ~ | J(x_d, t') | }  .
\end{eqnarray}
Here, we use this formalism to see how arrival times are affected by the intrinsic decoherence. 
For a given state, the probability current density is positive for all times at the location of the detector 
which, by assumption, is taken to be at the origin. Then, the normalized arrival time distribution is
\begin{eqnarray} \label{eq: ardis}
\Pi(0, t) &=& \frac{ J(0, t)  }{ \int_0^{\infty}dt' ~ J(0, t') }  .
\end{eqnarray}
The mean arrival time at the detector location $ x_d = 0 $ and its variance  are then given by
\begin{eqnarray}
\la t \ra &=& \int_0^{\infty}dt' ~ t' ~ \Pi(0, t')   , \label{eq: meanar} \\
(\Delta t)^2 &=& \la t^2 \ra - \la t \ra^2 , \label{eq: varar} 
\end{eqnarray}
respectively.
The standard deviation and the square root of the variance can be used as a measure of the width of the arrival 
time distribution, i.e., the rms width. Then, from Eqs. (\ref{eq: mil_sol}) and (\ref{eq: mil-jd}), we have that 
\begin{eqnarray} 
\int_0^{\infty}dt' ~ J(0, t') &=& \frac{1}{2\pi i} \int_{-\infty}^{\infty} du \int_{-\infty}^{\infty} dv \frac{\rho(u, v, 0)}{v} , \label{eq: j_tint} \\
\int_0^{\infty}dt' ~ t'~J(0, t') &=& -\frac{m \hb}{2\pi} \int_{-\infty}^{\infty} du \int_{-\infty}^{\infty} dv \frac{\rho(u, v, 0)}{ u v^2 \left( 1- \frac{i u v}{2m \hb \ga} \right) } , \label{eq: j*t_tint} \\
\int_0^{\infty}dt' ~t'^2~ J(0, t') &=& - \frac{m^2\hb^2}{i \pi} \int_{-\infty}^{\infty} du \int_{-\infty}^{\infty} dv \frac{\rho(u, v, 0)}{  u^2 v^3 \left( 1- \frac{i u v}{2m\hb\ga} \right)^3 }    , \label{eq: j*t^2_tint}
\end{eqnarray}
where we have used the new variables
\begin{numcases} ~
u = \frac{p+p'}{2} \\
v = p - p'   .
\end{numcases}
Eq. (\ref{eq: j_tint}) shows that the normalization factor (or time integral of this current) is independent 
on $ \ga^{-1} $ i.e., the intrinsic decoherence does not affect this quantity. The first and second moment 
have the same functional dependence of the intrinsic decoherence. From Eqs. (\ref{eq: j_tint}), (\ref{eq: j*t_tint}) and 
(\ref{eq: j*t^2_tint}), one writes to first order in  $ \ga^{-1} $
\begin{eqnarray}
\la t \ra &\approx & \la t \ra_{\QM} + \frac{1}{2\ga} \label{eq: tar_approx} \\
(\Delta t)^2 &\approx & (\Delta t)^2_{\QM} + \frac{2}{\ga} \la t \ra_{\QM}    , \label{eq: varar_approx}
\end{eqnarray}
where the sub-index ``q'' stands for the corresponding quantity within the standard quantum mechanics. 
Eqs. (\ref{eq: tar_approx}) and (\ref{eq: varar_approx}) confirm that both the mean arrival time and the rms 
width of the arrival time distribution are increased by the intrinsic decoherence and the amount of increment 
is linear in $ \ga^{-1} $. We emphasize that the above results are only valid for states where the current
density is positive.
It is noteable that a master equation in the same form of the Milburn equation has been derived to study errors in measurement of time by a clock \cite{Maetal-LNP-2008}. Treating the arrival time model as a kind of clock, Eqs. \eqref{eq: tar_approx} and \eqref{eq: varar_approx} can be interpreted as a limit on clock precision.

\subsection{ Transmission probability through a localized barrier}

Consider now the solution of the time independent Schr\"odinger equation for a localized barrier $V(x)$,
\begin{eqnarray} \label{eq: tise-sol}
u_k(x) &=& \frac{1}{\sqrt{2\pi}}
\begin{cases}
e^{i k x} + R(k) e^{-ikx} & x < 0 , \\
A(k) f(x) + B(k) g(x) &  0 \leq x < L  , \\
T(k) e^{ikx} & L \leq x   ,
\end{cases}
\end{eqnarray}
where $ k = \sqrt{2mE/\hb^2} $ is the wave number, $ [0, L] $ is the support of the barrier and 
$R(k)$ and $T(k)$ are the reflection and transmission coefficients, respectively. When $ E < V_{\max} $, 
tunneling is present. Finally, $ f(x) $ and $ g(x) $ are two independent solutions of the Scr\"odinger equation
in the barrier region. Note that here we have preferred to use $ u_k(x) $ instead of $ u_E(x) $.

Let us now consider transmission of a localized wave packet through the barrier. If the system is initially 
prepared in the state $ \psi_0(x) $ in the far left side of the barrier and the wavepacket includes only 
positive momenta, the solution of the Schr\"odinger equation is given by 
\begin{eqnarray} \label{eq: tdse-sol}
\psi(x, t) &=& \int_0^{\infty} dk ~ \phi(k) e^{-i E_k t/ \hb} u_k(x), \qquad E_k = \frac{\hb^2 k^2}{2m}   ,
\end{eqnarray}
where $ \phi(k) = \la k | \psi_0 \ra $ denotes the momentum amplitude (the Fourier transform of $\psi_0(x)$) 
of initial wave packet in the absence of the barrier.

From Eq. \eqref{eq: rhot_EE'}, the solution of the corresponding problem in the Milburn framework is given by
\begin{eqnarray} \label{eq: rhotEE'-barrier}
\rho(x, x',t) &=&  \int_0^{\infty} dk ~ \int_0^{\infty} dk' ~ 
e^{-i(E_k -E_{k'})t/\hb - (E_k -E_{k'})^2 t / (2\hb^2 \ga)  } \phi(k) \phi^*(k') u_k(x) u_{k'}^*(x')  .
\end{eqnarray}
%
%
%
The time-dependent transmission probability is defined as
\begin{eqnarray} \label{eq: trans-prob}
P_{\tr}(t) &=& \int_L^{\infty} dx ~ \rho(x, x, t)
\end{eqnarray}
and the stationary value of this quantity is independent on the lower limit of the corresponding integral. 
The transmission probability for a monochromatic beam with momentum $ p=\hb k $ is $ |T(k)|^2 $. 
Due to the fact that the initial probability density in momentum is $ |\phi(k)|^2 $, then the stationary 
transmission probability for a wave packet is
\begin{eqnarray} \label{eq: trprob-st}
P_{\tr}(\infty) &=& \int_0^{\infty} dk ~ |\phi(k)|^2 |T(k)|^2  .
\end{eqnarray}
This is applicable even in the framework of the Milburn equation. 
In this way, one sees that $ P_{\tr}(\infty) $ is independent on the intrinsic decoherence parameter $\ga$.

\section{ Numerical results and discussions }

In this section, several problems are analyzed numerically within the Milburn context in order to see how the 
intrinsic decoherence process is settled down in the dynamics. We are going to work in a system of 
units where $ \hb = m = 1 $.

\subsection{Interference of two wave packets}

As a first application, we analyze the effect of the intrinsic decoherence on the interference of two wave packets in free space i.e., in the absence of external interactions. Consider the superposition 
\begin{eqnarray} \label{eq: Gauss_sup}
\psi(x, 0) &=& \mathcal{N} ( \psi_a(x, 0) + \psi_b(x, 0) )  
\end{eqnarray}
of two well-separated Gaussian wave packets, each one described by Eq. \eqref{eq: Gauss} with the same 
width $\si_0$, well-localized at $ x_{0a} $ and $ x_{0b} $ and  kick momenta $ p_{0a} $ and $ p_{0b} $. 
With these initial conditions,
\begin{numcases}~
x_{0a} = - x_{0b} = x_0  , \label{eq: x0ax0b}\\
p_{0a} = - p_{0b} = p_0 , \label{eq: p0ap0b}
\end{numcases}
the normalization constant $ \mathcal{N} $ is given by
\begin{eqnarray} \label{eq: norcon}
 \mathcal{N} &=& \left \{ 2 \left( 1 + \exp \left[ - \frac{x_0^2}{2\si_0^2} - 2 \frac{p_0^2 \si_0^2}{\hb^2} \right] \right) \right \}^{-1/2} .
\end{eqnarray}
The time evolution is given by Eq. (\ref{eq: rho_pos_mil}) and the corresponding probability density reads as
\begin{eqnarray} \label{eq: pd-pos}
P(x, t) &=& \rho(R=x, r =0, t) =  \mathcal{N}^2 [ P_{aa}(x, t) + P_{bb}(x, t) + 2 \re \{P_{ab}(x, t)\} ]   ,
\end{eqnarray}
where $ P_{ij}(x, t) = \varrho_{ij}(R=x, r =0, t) $ give the diagonal elements of the time evolution of 
$ \psi_i(x, 0) \psi_j^*(x', 0) $
\begin{eqnarray} \label{eq: rho-pos_ij}
P_{ij}(x, t) &=& \frac{1}{2\pi \hb} \int_{-\infty}^{\infty}dp  \int_{-\infty}^{\infty}dp'
~ e^{i(p-p')x/\hb} \exp \left[ - \frac{i}{\hb} \frac{p^2-p'^2}{2m} t - \frac{1}{2\hb^2 \ga} \left( \frac{p^2-p'^2}{2m} \right)^2  t \right]
\nonumber \\
& \times & \sqrt{\frac{2}{\pi}} \frac{\si_0}{\hb}
\exp \left[ - \frac{ ( p - p_{0i} )^2 + ( p' - p_{0j} )^2  }{ \hb^2 } - i \frac{ x_{0i} ~ p  - x_{0j} ~p' }{\hb} \right]  .
\end{eqnarray}
%
%
%
\begin{figure}  
	\includegraphics[width=14cm,angle=-0]{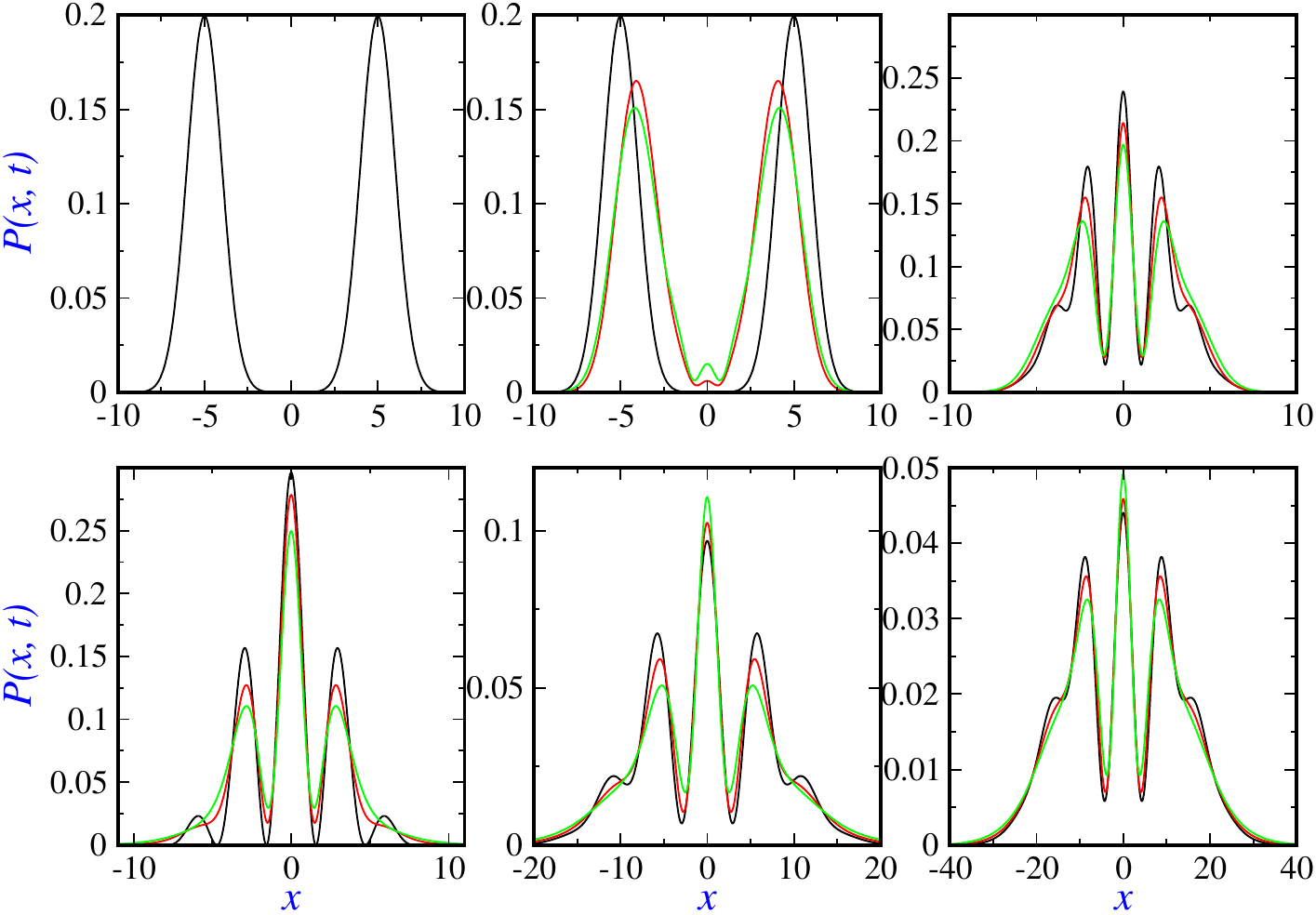}
	\caption{
		The probability density for the cat state \eqref{eq: Gauss_sup} in free propagation under the Milburn equation 
		versus position at different times $ t = 0 $ (left top panel) $ t = 1 $ (middle top panel),  $ t = 3 $ (right top panel), 
		$ t = 5 $ (left bottom panel), $ t = 10 $ (middle bottom panel),  $ t = 15 $ (right bottom panel)
		for $ \ga^{-1} = 0 $ (black curves), $ \ga^{-1} = 0.2 $ (red curves) and $ \ga^{-1} = 0.5 $ (green curves).	
		For numerical calculations, we have used $ \si_0 = 1 $, $ x_0 = 5 $, $ p_0 = 1 $, $ x_{0a} = - x_{0b} = x_0 $ 
		and $ p_{0a} = - p_{0b} = - p_0 $.
	}
	\label{fig: interference}
\end{figure}
%
%
%
\begin{figure}  
	\includegraphics[width=12cm,angle=-0]{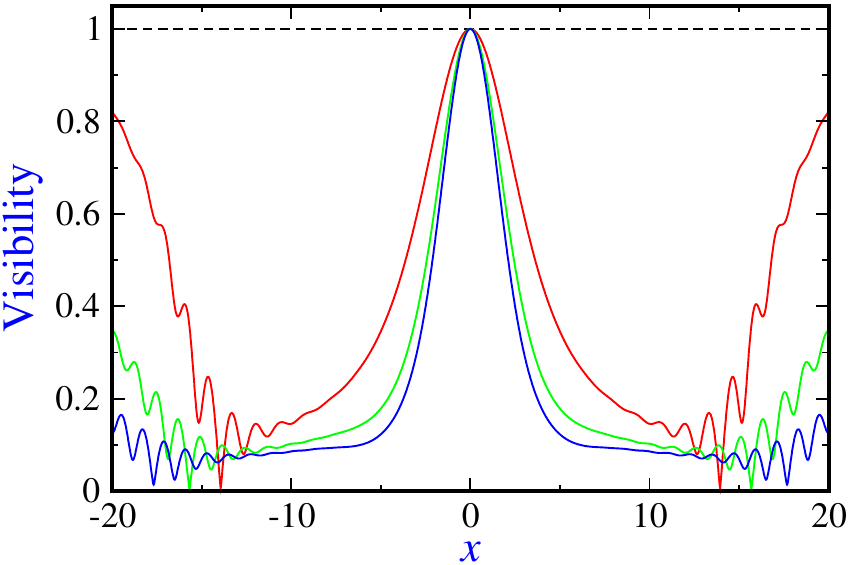}
	\caption{
		The visibility function $ \mathcal{V}(x, t_0) $ versus $x$ at time $ t_0 = m x_0/p_0 = 5 $ for different values 
		of the intrinsic decoherence parameter: for $ \ga^{-1} = 0 $ (black, dashed curve),$ \ga^{-1} = 0.2 $ (red curve), 
		$ \ga^{-1} = 0.5 $ (green curve) and $ \ga^{-1} = 0.8 $ (blue curve). The same parameters are used as 
		Figure \ref{fig: interference}.	}
	\label{fig: visibility}
\end{figure}
%
In Figure \ref{fig: interference}, the probability density (\ref{eq: pd-pos}) at different times for different values of the intrinsic decoherence parameter is plotted. As this figure shows, interference fringes are apparent for the 
von Neumann dynamics while disappearing with increasing $ \ga^{-1} $. 
When particles propagate under the usual von Neumann equation i.e., for $ \ga^{-1} = 0 $, the 
Gaussian wave packets $ \psi_a $ and $ \psi_b $ with the conditions given by Eqs. (\ref{eq: x0ax0b}) and 
(\ref{eq: p0ap0b}) overlap at  time $ t_0 = m x_0 / p_0 $. For these initial conditions, from Eq. (\ref{eq: rho-pos_ij}) 
and for $ \ga^{-1} = 0 $ one can easily show that
\begin{eqnarray} \label{eq: schprobs}
P_{aa}(x, t_0) &=& P_{bb}(x, t_0) = | P_{ab}(x, t_0) | = \frac{1}{ \sqrt{2\pi} \si_{t_0}}
\exp \left[ - \frac{2 p_0^2 \si_0^2 }{\hb^2 x_0^2 + 4 p_0^2 \si_0^4} x^2 \right], \\
 \si_t &=& \si_0 \sqrt{ 1 + \frac{\hb^2 t^2}{4 m^2 \si_0^4} }   .
\end{eqnarray}
Following \cite{Mi-PRA-1991}, one can define the fringe visibility as
\begin{eqnarray} \label{eq: visibility}
\mathcal{V} &=& \frac{ | P_{ab}(x, t_0) | }{ \sqrt{ P_{aa}(x, t_0) P_{bb}(x, t_0) } }  ,
\end{eqnarray}
which is a way to quantify the degree of interference. 
From Eqs. \eqref{eq: schprobs} and \eqref{eq: visibility}, one sees that the visibility equals unity, independent on $x$.
Figure \ref{fig: visibility} shows the visibility for different values of  $ \ga^{-1} $. According to this figure, and as 
one expects, around the central peak, the visibility in the Milburn framework is a decaying function of $ |x| $ and 
is smaller for higher values of $ \ga^{-1} $.

\subsection{Tunneling through a rectangular barrier}

We now investigate tunneling through a rectangular barrier
\begin{eqnarray} \label{eq: bar-pot}
	V(x) &=& V_0 \theta(x) \theta(L - x)   .
\end{eqnarray}
Let us consider the Gaussian wave packet
\begin{eqnarray} \label{eq: Gauss}
	\psi(x) &=& \frac{1}{ (2\pi \si_0^2)^{1/4} } \exp \left[ - \frac{x-x_0}{4\si_0^2} + i \frac{ p_0 }{\hb} x \right]  ,
\end{eqnarray}
which is incident from left. 
Parameters of the wave packet and the barrier have been chosen in a way that: (i) contribution of negative 
momenta to the wave packet is negligible and (ii) the mean energy of the wave packet is smaller than the height 
of the barrier. Thus, transmission is present due to tunneling. In particular, we use: $ \si_0=1 $, $ p_0=2 $, 
$ x_0= -10 $, $V_0=3$ and $L=1$. 
%

%
\begin{figure}  
	\includegraphics[width=12cm,angle=-0]{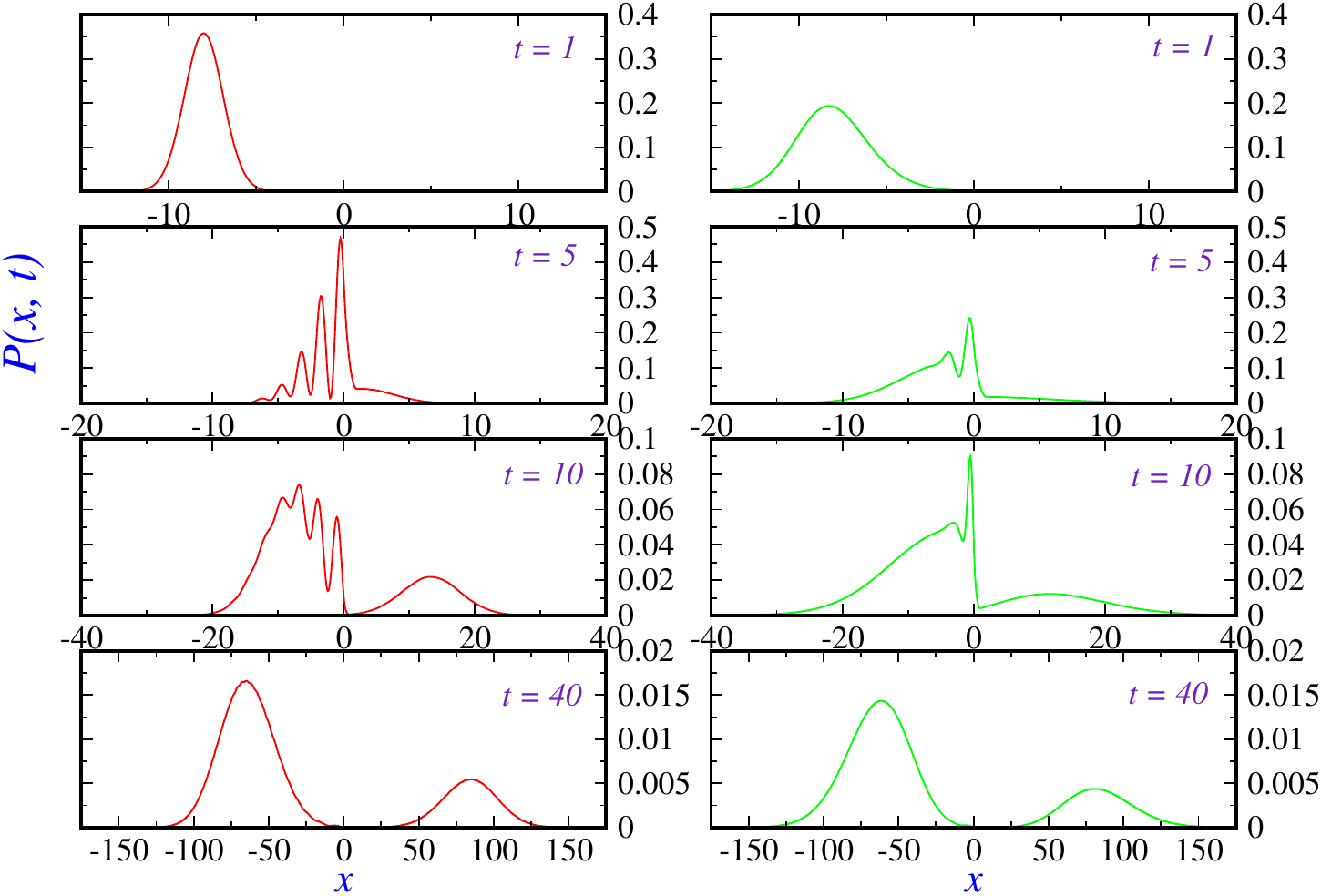}
	\caption{
		Probability density for incident of a Gaussian wave packet on a rectangular barrier for the von Neumann dynamics 
		(left column) and the Milburn dynamics with $ \ga^{-1} = 0.8 $ (right column) at different times: $ t =1 $ (first row), 
		$ t = 5 $ (second row), $ t = 10 $ (third row) and $ t = 40 $ (fourth row). 
		For numerical calculations we have used: $ \hb = m = 1 $, $ \si_0=1 $, $ p_0=2 $, $ x_0= -10 $, $V_0=3$ and $L=1$. }
	\label{fig: rhoSchMil}
\end{figure}
%

\begin{figure}  
	\includegraphics[width=12cm,angle=-0]{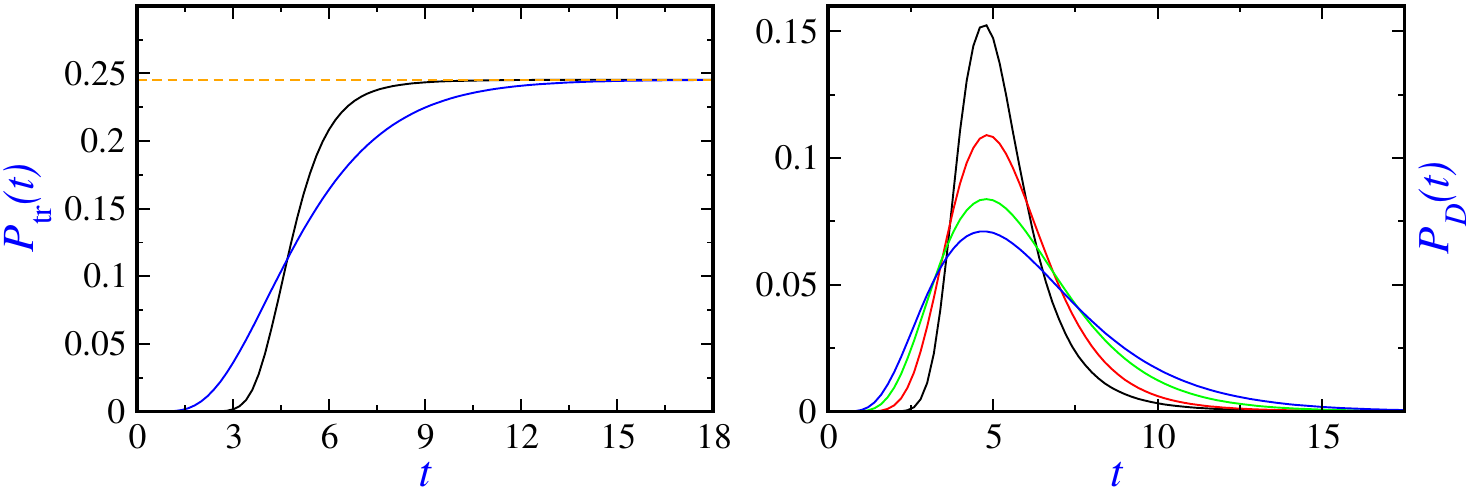}
	\caption{
		Time dependent transmission probability (left panel) and probability inside the rectangular barrier region 
		(right panel) for an incident  Gaussian wave packet for $ \ga^{-1} = 0 $ (black curves),$ \ga^{-1} = 0.2 $ 
		(red curve), $ \ga^{-1} = 0.5 $ (green curve) and $ \ga^{-1} = 0.8 $ (blue curves). 
		Orange dashed line on the left panel depicts the stationary value of the transmission probability which 
		is the same for both,  the von Neumann and Milburn dynamics.
		The same parameters as Figure \ref{fig: rhoSchMil} have been used.}
	\label{fig: TrprobDwell}
\end{figure}

In Figure \ref{fig: rhoSchMil}, the probability density is plotted at different times for the von Neumann 
(left column) and  Milburn (right column) dynamics. It is clearly seen that the reflection and transmission 
parts of the wave packet are quite reduced for $ \ga^{-1} = 0.8 $. The oscillations of the reflection part of 
the wave packet are also nearly suppressed. This behavior is better understood from Figure 
\ref{fig: TrprobDwell} where both the transmission probability and the probability inside the barrier region 
are depicted for both dynamics. As calculations confirm, the stationary value of the transmission probability 
and the mean dwell time given by
\begin{eqnarray} \label{eq: tauD}
	\tau_{\text{D}} &=& \int_0^{\infty} dt~ P_{\text{D}}(t)
\end{eqnarray}
are the same for both dynamics where 
\begin{eqnarray} \label{eq: PrD}
	P_{\text{D}}(t) &=& \int_0^L dx ~ \rho(x, x, t)
\end{eqnarray}
is the probability for finding the particles inside the barrier region. However, note that in the Milburn dynamics, 
particles arrive at the barrier sooner while 
leave later. For these chosen parameters, the stationary transmission probability is $ 0.24536 $; independent on 
the value of the intrinsic decoherence parameter $ \ga $ and the mean dwell time is $ 0.4184 $, again the same for both dynamics.

One simple explanation for the independence of the dwell time from $ \ga^{-1} $ is the following. As the stationary solutions of the Milburn equation are the same as those of the von Neumann one, then 
the average dwell time for a monochromatic beam with a given momentum \cite{Muetal-LNP-2009, Bu-PRB-1983} is the same for both, i.e.,
\begin{eqnarray}
\tau_{\text{D}}(k) &=& \frac{1}{|j(k)|} \int_0^L dx |u_k(x)|^2  ,
\end{eqnarray}
where $ j(k) = \frac{1}{2\pi} \frac{\hb k}{m} $ is the incident flux corresponding to the incident wave function $ e^{i k x} / \sqrt{2\pi} $. Thus, the dwell time associated with the incoming state $ \hat{\rho}(0) $ i.e.,
\begin{eqnarray}
\tau_{\text{D}} &=& \int_0^\infty dk ~ \la k | \hat{\rho}(0) | k \ra ~ \tau_{\text{D}}(k)
\end{eqnarray}
is also the same for both dynamics.

%
%

\subsection{The bouncing ball}

\begin{figure}  
\includegraphics[width=10cm,angle=-0]{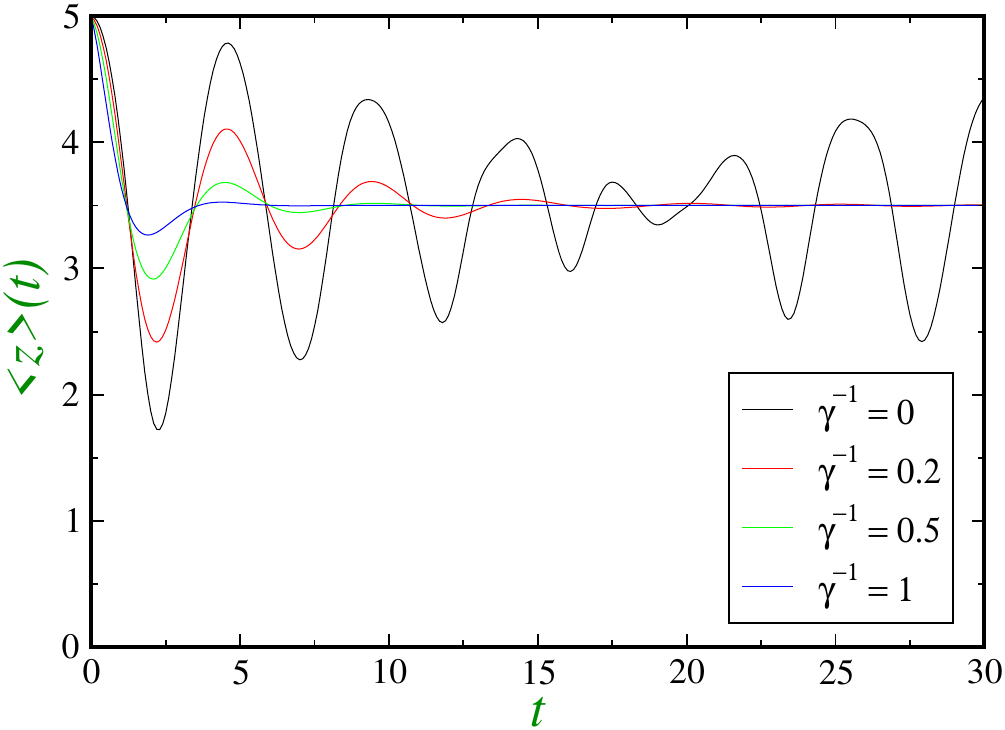}
\caption{
Position expectation value $ \la z \ra(t) $ for the Milburn dynamics. For numerical calculations we have used 
$\si_0 = 1$ and $ z_0 = 5 $. Space coordinates in units of $ (\hb^2/2m^2g)^{1/3} $ and time in units of 
$ (2\hb/mg^2)^{1/3} $.}
\label{fig: zav_mil}
\end{figure}

As the last application of the Milburn formalism, we analyze a bound system with discrete energy spectrum, the 
bouncing ball. This ball has mass $m$ and is bouncing vertically over a table in the presence of the gravity 
field $g$,
\begin{eqnarray} \label{eq: pot}
V(z) &=& 
\begin{cases}
m g z & z > 0   ,\\
+\infty & z \leq 0  .
\end{cases}
\end{eqnarray}
Here, we have denoted the space coordinate by $z$ instead of $x$ in order to emphasize that motion is in 
the vertical direction. The solution of the time-independent Schr\"odinger equation 
\begin{eqnarray} \label{eq: TISE}
\left( \frac{\hb^2}{2m} \frac{d^2}{d z^2} + m g z \right) u_n(z) &=& E_n u_n(z), \qquad z > 0
\end{eqnarray}
yields for the eigenvalues and eigenfunctions \cite{VaSo-book-2004}
\begin{numcases}~
E_n = - R_n \left( \frac{1}{2} m g^2 \hb^2 \right)^{1/3} \label{eq: En-exact} \\
u_n(z) = \sqrt{\al} \frac{1}{\Ai'(R_n)} \Ai(\al z + R_n)   , \qquad \al = \left( \frac{2 m^2 g}{\hb^2} \right)^{1/3}  , 
\label{eq: eigenfunc}
\end{numcases}
respectively. Here $\Ai(z)$ and $\Ai'(z)$ are the Airy function and its derivative, $R_n$ is the n$^{\text{th}}$ root 
of the Airy function i.e., $ \Ai(R_n) = 0 $. Let the initial wave function be a Gaussian wave packet 
\eqref{eq: Gauss} with $ p_0 = 0 $. In this case, the  expansion coefficients (\ref{eq: expan-coeff}) are given by
\begin{eqnarray} \label{eq: expan-bb}
C_n &=& \frac{ \sqrt{\al} }{\Ai'(R_n)} \frac{1}{(2\pi \si_0^2)^{1/4}} \int_0^{\infty} dz ~ \Ai(\al z + R_n) \exp \left[ - \frac{(z-z_0)^2}{4\si_0^2} \right]   .
\end{eqnarray}
%
%
\begin{figure}  
	\includegraphics[width=12cm,angle=-0]{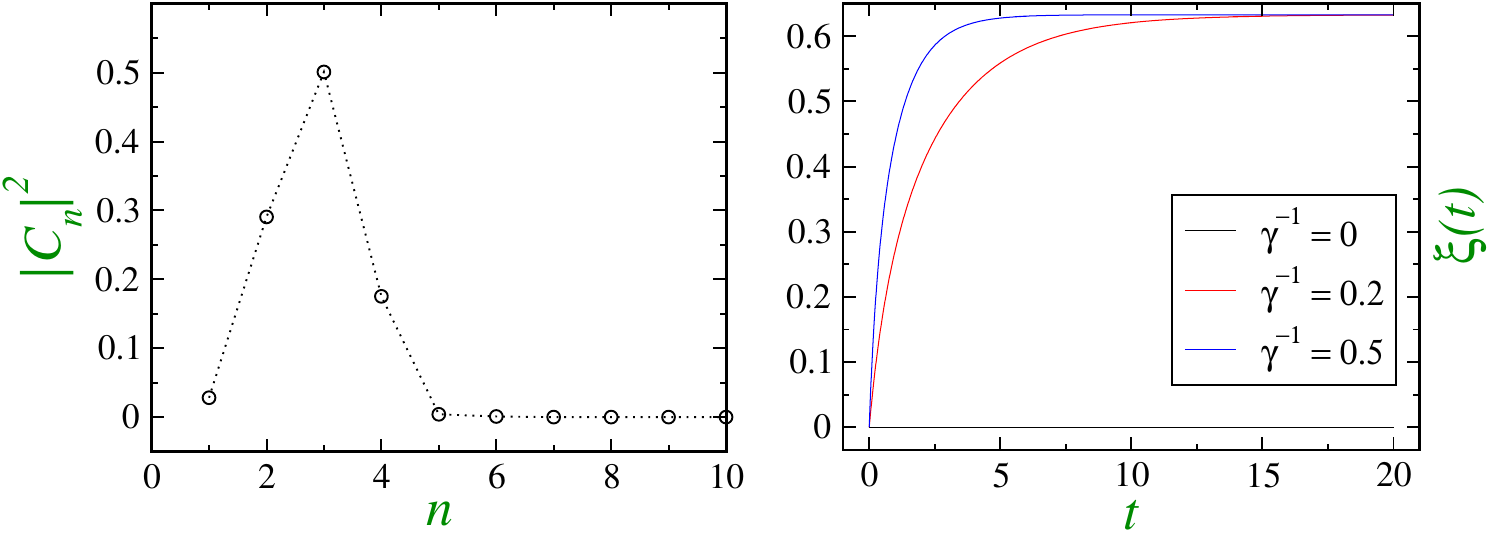}
	\caption{
		Expansion weights $ |C_n|^2 $ versus $n$ (left panel) computed by Eq. (\ref{eq: expan-bb}) and the linear entropy \eqref{eq: pu_def}
		versus time for different values of $ \ga^{-1}$, in units of $ (2\hb/mg^2)^{1/3} $.
		For numerical calculations, we have used $\si_0 = 1$ and $ z_0 = 5 $ in units of $ (\hb^2/2m^2g)^{1/3} $.}
	\label{fig: coef-pur}
\end{figure}
%
From Eq. (\ref{eq: prob-lim}), one obtains
\begin{eqnarray} \label{eq: zav-lim}
\lim_{t \to \infty} \la z \ra(t)  & \approx & \sum_n |C_n|^2 \int dz~z |u_n(z)|^2  
= \sum_n |C_n|^2 \la u_n | z | u_n \ra
\end{eqnarray}
for the asymptotic value of the position expectation value.
By substituting Eq. (\ref{eq: eigenfunc}) into (\ref{eq: zav-lim}), one reaches \cite{Va-AJP-2000}
\begin{eqnarray} \label{eq: zav-longtime}
\lim_{t \to \infty} \la z \ra(t)  & \approx & - \frac{2}{3 \al} \sum_n |C_n|^2 R_n  .
\end{eqnarray}
On left panel of Figure \ref{fig: coef-pur}, we have plotted the weights $ |C_n|^2 $ versus $ n $ for  
$\si_0 = 1$ and $ z_0 = 5 $ in units of $ (\hb^2/2m^2g)^{1/3} $ as the natural unit of length. Furthermore, 
natural units for energy and time are $ (mg^2 \hb^2/2)^{1/3} $ and $ (2\hb/mg^2)^{1/3} $, respectively.
As this figure shows, the first few terms are enough for calculations. The right panel of this figure depicts the 
evolution of linear entropy for different values of $\ga^{-1}$. 
Concerning the bouncing motion, Figure \ref{fig: zav_mil} shows that oscillations are killed by increasing 
the Milburn parameter $ \ga^{-1} $. Taking only the first ten expansion coefficients for the probability density, 
one has that  $ \lim_{t \to \infty} \la z \ra(t)  \approx 3.5 $ from Eq. (\ref{eq: zav-longtime}) which coincides with 
Figure \ref{fig: zav_mil}. 

%
%

\section{Conclusions}

In this work, we have analyzed several theoretical and practical aspects of the Milburn equation which comes from 
a first order expansion on the  parameter $\gamma^{-1}$ governing the decoherence, the so-called intrinsic 
decoherence.  In the limit of  $ \gamma^{-1} \to 0 $, one recovers the von Neumann equation for the 
density operator. Linear entropy is reduced following a linear behavior in time. Ehrenfest relations and the 
probability current density and arrival times are also modified by the presence of the intrinsic decoherence. 
In particular, the extension of the Milburn formalism to include the Wigner representation of the 
corresponding equation and the relation to a Lindbladian master equation are important steps to better
understand the importance of the corresponding quantum dynamics. 
Moreover, the interference displayed by two Gaussian wave packets leads to a decreasing
function of the visibility as the intrinsic decoherence increases. The same is observed for the tunneling
transmission probability. For the bouncing ball problem,  the oscillations displayed by the average value of 
the height are clearly attenuated. In a certain sense, this behavior is the one should expect when the 
external decoherence is present; a gradual and continuous  process. This work stresses thus new insights
in the Milburn dynamics compared to the von Neumann dynamics of the density operator.

\begin{acknowledgements}

SVM acknowledges support from the University of Qom. SMA acknowledges support of a grant from 
the Ministery of Science, Innovation and Universities with Ref. PID2023-149406NB-I00.

\end{acknowledgements}

\vspace{0.5cm}

{\bf Data availability}: This manuscript has no associated data.



%
\end{document}